# Sensitivity of Cosmological Parameter Estimation to Nonlinear Prescription from Cosmic Shear

Sarah Safi and Marzieh Farhang
Department of Physics, Shahid Beheshti University, 1983969411, Tehran, Iran

## Abstract

Several ongoing and upcoming large-scale structure surveys aim to explore the nonlinear regime of structure formation with high precision. Making reliable cosmological inferences from these observations necessitates precise theoretical modeling of the mildly nonlinear regime. In this work we explore how the choice of nonlinear prescription would impact parameter estimation from cosmic shear measurements for a Euclid-like survey. Specifically, we employ two different nonlinear prescriptions of halofit and the Effective Field Theory of the Large Scale Structure and compare their measurements for the three different cosmological scenarios of $\Lambda$CDM, $w$CDM, and $(w_0, w_a)$ CDM. We also investigate the impact of different nonlinear cutoff schemes on parameter estimation. We find that the predicted errors on most parameters shrink considerably as smaller scales are included in the analysis, with the amount depending on the nonlinear prescription and the cutoff scheme used. We use predictions from the halofit model to analyze the mock data from DarkSky $N$-body simulations and quantify the parameter bias introduced in the measurements due to the choice of nonlinear prescription. We observe that $\sigma_8$ and $n_s$ have the largest measurement bias induced by inaccuracies of the halofit model.

*Unified Astronomy Thesaurus concepts:* Cosmology (343); Large-scale structure of the universe (902); Observational cosmology (1146); Astronomy data analysis (1858); Weak gravitational lensing (1797)

## 1. Introduction

Current and forthcoming large-scale structure surveys are expected to make remarkable improvements in accurately probing the mildly nonlinear regime (MNL) of the structure formation in the Universe (see, e.g., Laureijs et al. 2011; Ivezić et al. 2019; Bailey et al. 2023), thereby contributing to higher precision of parameter estimation. Within the large volumes of data, a significant amount of information is encapsulated in small scales. Considerable research efforts have been dedicated to investigating this regime, aimed to achieve enhanced cosmological inferences (e.g., Martinelli et al. 2011, 2020; Wang 2012; Audren et al. 2013; Bull 2016; Sartoris et al. 2016; Abbott et al. 2018; Sprenger et al. 2019; Safi & Farhang 2021; Fumagalli et al. 2024).

When managing these data sets, the issue of dealing with nonlinear behavior becomes an important challenge. Therefore, there is a demand for theoretical modelings that effectively, yet precisely, describe the MNL regime. Various perturbative approaches have been proposed to investigate this regime such as standard perturbation theory (Bernardeau et al. 2002; Carlson et al. 2009), renormalized perturbation theory (Crocce & Scoccimarro 2006), regularized perturbation theory (Taruya et al. 2012), and the effective field theory of large-scale structure or EFTofLSS (Carrasco et al. 2014). A list of theoretical methods, with the specific scales for which their predictions hold, is presented in Carlson et al. (2009).

This study is a continuation of our previous research, where we analyzed the impact of the choice of the nonlinear prescription on cosmological parameter measurements obtained from galaxy clustering surveys, up to $k_{\max} = 0.3h$ Mpc$^{-1}$ (Safi & Farhang 2021). In this work we use simulations of cosmic shear data for a Euclid-like survey and explore the impact of the prescription for modeling the MNL regime on parameter estimation. In particular, we investigate how inaccuracies in the nonlinear prescription lead to biases in parameter evaluation. We also exploit the available information in these regimes to constrain parameters in a couple of different cosmological scenarios. As the nonlinear prescription, we use EFTofLSS, which is claimed to offer a convergent perturbation theory (Baumann et al. 2012; Carrasco et al. 2012, 2014). The model considers the impact of short-wavelength modes on large scales through introducing certain free coefficients to be measured by data. We compare its predictions with those of halofit (HF), a widely used approach for modeling the nonlinear regime in cosmology (Smith et al. 2003; Takahashi et al. 2012).

The rest of this paper is organized as follows. In Section 2, we describe in detail the theoretical modeling of the cosmic shear power spectrum and the mock data used in this work. In Section 3, we introduce the cosmic shear likelihood and the corresponding Fisher matrix. The results obtained from this analysis are then presented in Section 4. We conclude in Section 5.

## 2. Simulation of Cosmic Shear Power Spectrum

The recipe for modeling the structures in the MNL regime must be accurate enough to properly describe the high-resolution data from future surveys and make reliable forecasts. This section is dedicated to an explanation of the simulated data utilized for the analysis followed by introducing the nonlinear recipes and the dark energy models used in this study.

### 2.1. Simulations of Cosmic Shear

In a cosmic shear survey, the shapes and orientations of distant galaxies are measured to study the large-scale structure of the Universe. This is achieved by mapping the galaxy alignments due to lensing by the gravitational pull of the intervening structures, such as galaxies and clusters. Scientists use auto- and cross-correlations of these alignment maps at various redshifts to extract cosmological information. (Kilbinger 2015). The cosmic shear power spectrum in the Limber approximation is given by (Blanchard et al. 2020)

$$P_{ij}(\ell) \simeq \frac{9H_0^4}{4c^3} \int dz \frac{W_i(z)W_j(z)(\Omega_m(1+z))^2}{H(z)} \times P_m\left[\frac{\ell}{r(z)}, z\right], \quad (1)$$

where $i$ and $j$ represent the observed redshift bins, $\Omega_m$ is the current density parameter for the matter, $P_m(k, z)$ is the matter power spectrum at $k = \ell/r(z)$, and $W_i$ is the weight function for the $i$th bin, described by

$$W_i(z) = \int_z^{z_{max}} dz' \, n_i(z') \left[1 - \frac{r(z)}{r(z')}\right]. \quad (2)$$

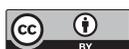

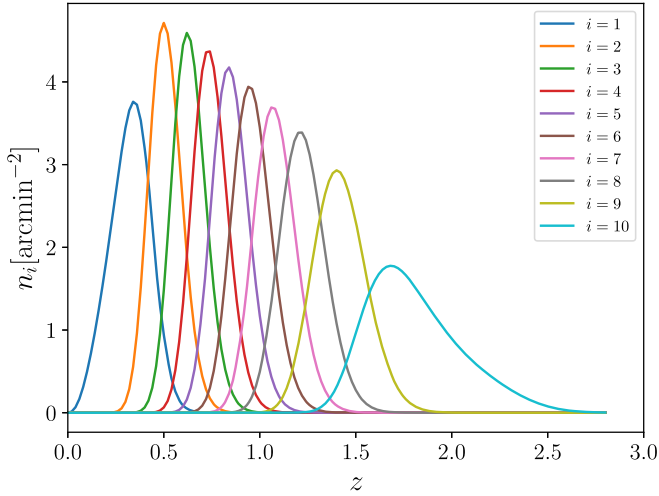

**Figure 1.** Galaxy redshift distribution for a Euclid-like survey (Equation (3)).

$n_i(z)$ is the number density of the *observed* galaxies in the $i$th bin that can be written as

$$n_i(z) = \frac{\int_{z_i^-}^{z_i^+} dz_p \, n(z) p_{ph}(z_p|z)}{\int_{z_{min}}^{z_{max}} dz \int_{z_i^-}^{z_i^+} dz_p \, n(z) p_{ph}(z_p|z)}, \quad (3)$$

with ($z_i^-$, $z_i^+$) representing the bin edges, and $z_{min}$ and $z_{max}$ are the redshift limits of the survey. The denominator is chosen to normalize $n_i(z)$, such that $\int_{z_{min}}^{z_{max}} n_i(z) dz = 1$. The underlying *true* distribution $n(z)$ appearing in this expression is given by

$$n(z) \propto \left(\frac{z}{z_0}\right)^2 \exp\left[-\left(\frac{z}{z_0}\right)^{3/2}\right], \quad (4)$$

where $z_0 = z_m/\sqrt{2}$, with $z_m = 0.9$ being the median redshift. The distribution function $p_{ph}(z_p|z)$ describes the probability that a galaxy truly located at $z$ is observed at $z_p$, and it is assumed to be Gaussian. Figure 1 illustrates the galaxy redshift distribution for the redshift bins of a Euclid-like survey (see Table 1).

The overall impact of the bin weight functions on the cross power spectra, $P_{ij}(\ell)$, is determined by $W_iW_j$ and is illustrated for several bin pairs in Figure 2. We see that, as expected, $W_iW_j$ drops to an insignificant level at high $z$ for higher redshift bins, therefore considerably reducing the contribution of the matter power spectra $P_m$ at those $z$ to $P_{ij}(\ell)$. The cosmic shear probe of Euclid is planned to observe sources between redshifts of 0 and 2.5, in ten equipopulated tomographic redshift bins with edges given in Table 1 (Euclid Collaboration et al. 2023).

In addition to the lensing-induced alignment, the galaxies also have some intrinsic alignments. The uncorrelated part of this alignment is modeled to contribute to the total power spectrum as shot noise. We thus have

$$C_{ij}(\ell) = P_{ij}(\ell) + N_{ij}, \quad (5)$$

where

$$N_{ij} = \frac{\sigma_\epsilon^2}{\bar{n}_g}\delta_{ij} \quad (6)$$

is the shot noise term and $C_{ij}(\ell)$ is the total shear power spectrum. Here $\sigma_\epsilon^2$ is the variance of the observed ellipticities, the Kronecker delta guarantees that the shot noise is absent for the cross-correlation spectra as the ellipticities of galaxies at different redshifts are not expected to be correlated, and $\bar{n}_g$ is the surface density of galaxies per steradian for each bin, i.e.,

$$\bar{n}_g = \frac{n_{gal}}{N_{bin}} \times 3600\left(\frac{180}{\pi}\right)^2. \quad (7)$$

Table 2 presents the specifications for the Euclid survey (Laureijs et al. 2011; Martinelli et al. 2020).

### 2.2. Theoretical Modeling and Data

This section briefly introduces the nonlinear prescriptions and dark energy models applied in our study. The section also briefly presents the specifications of the DarkSky data set that was employed in this work. We generate simulations of the observed power spectrum of cosmic shear for two nonlinear prescriptions: HF and EFTofLSS. These simulations are carried out in accordance with the procedures outlined in Section 2.1, using the Euclid specifications outlined in Table 2.

#### 2.2.1. The Nonlinear Prescription

For a thorough analysis of data obtained from a Euclid-like experiment, it is crucial to have precise modeling of the matter power spectrum at a high level of accuracy, aiming for approximately 1% precision. HF is among the commonly used to describe the matter power spectrum in the MNL regime, which is based on fitting formulas derived from the results of $N$-body simulations (Ma & Fry 2000; Seljak 2000; Cooray & Sheth 2002). However, HF is claimed to have an accuracy of around 3%, making it unsuitable for reliable analysis in the context of upcoming high-resolution large-scale surveys. Additionally, the HF model is primarily calibrated for the $\Lambda$CDM cosmological scenario and lacks appropriate extensions to accommodate general non-$\Lambda$CDM models (Amendola et al. 2018). The present study aims to investigate the impact of these inadequacies on cosmological inferences and compare its performance to the predictions of the (semi)analytical model of EFTofLSS.

Table 1
The Edges of the Ten Tomographic Redshift Bins Used in This Work with the Corresponding Bin Edges for Observations by a Euclid-like Survey (Euclid Collaboration et al. 2023)

| $i$ | 1 | 2 | 3 | 4 | 5 | 6 | 7 | 8 | 9 | 10 | 11 |
|---|---|---|---|---|---|---|---|---|---|---|---|
| $z$ | 0.001 | 0.42 | 0.56 | 0.68 | 0.79 | 0.90 | 1.02 | 1.15 | 1.32 | 1.58 | 2.50 |

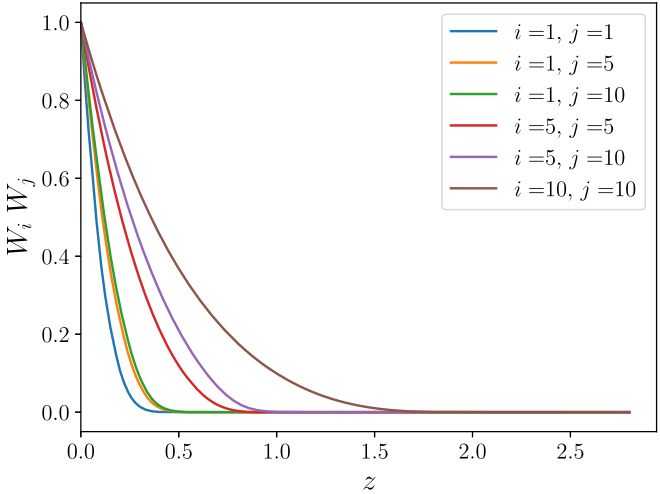

**Figure 2.** Lensing kernels for different bin pairs used in Equation (2).

Table 2
Euclid Specifications for Weak-lensing Survey

| | Parameter | Euclid |
|---|---|---|
| Survey area in the sky | $A_{\text{survey}}$ | 15,000 deg$^2$ |
| Galaxy number density | $n_{\text{gal}}$ | 30 arcmin$^{-2}$ |
| Total intrinsic ellipticity dispersion | $\sigma_\epsilon$ | 0.22 |
| Number of redshift bins | $N_{\text{bin}}$ | 10 |

EFTofLSS, referred to as EFT in the following discussion, offers an effective framework for describing the behavior of the Universe on large scales by integrating out the effects of short-wavelength perturbations. In this approach, long-wavelength modes are treated as an effective fluid characterized by a small number of parameters, including the equation of state, speed of sound, and viscosity. The coefficients governing the dynamics of the small-scale modes within EFT need to be determined by fitting to observational data or by measuring them through $N$-body simulations (Baumann et al. 2012; Carrasco et al. 2012).

EFT has been developed up to three-loop order, as demonstrated by previous works (Carrasco et al. 2014; Baldauf et al. 2015b; Konstandin et al. 2019). These developments include considerations for various modifications, such as bias and baryonic effects (Assassi et al. 2014; Angulo et al. 2015a; Lewandowski et al. 2015; Mirbabayi et al. 2015; Senatore 2015). Additionally, the theory's predictions have been compared to simulations in multiple aspects, encompassing the dark matter density power spectrum (Carrasco et al. 2012, 2014; Senatore & Zaldarriaga 2015), bispectrum (Angulo et al. 2015b; Baldauf et al. 2015a), dark matter momentum power spectrum (Senatore & Zaldarriaga 2015), and the dark matter power spectrum in redshift space (Senatore & Zaldarriaga 2014). The EFT predictions are claimed to have considerable improvement over the predictions from alternative models (Carrasco et al. 2012, 2014; Foreman et al. 2016).

In this work we adopt the parameterization introduced by Foreman et al. (2016) to efficiently encompass the effects of short-wavelength modes. We refer to these parameters collectively as sound speeds and use the generic label $c_s$ throughout the paper. The $c_s$ represents the four parameters $c_{s(1)}^2$s that characterize the one-loop counterterms in the nonlinear correction to the matter power spectrum (see Equation (6.1) of Foreman et al. 2016). Redshift dependence of the counterterms is parametrized with fitting functions by four free parameters (see Equation (5.7) of Foreman et al. 2016).

### 2.2.2. Models of Dark Energy

In this work we explore three sets of cosmological scenarios: the $\Lambda$CDM model, the $w$CDM model with a constant $w$ as the dark energy equation of state, which can deviate from $-1$, and a dynamical dark energy model with its equation of state described by (Chevallier & Polarski 2001; Linder 2003)

$$w(a) = w_0 + (1-a)w_a. \quad (8)$$

Here, $w_0$ and $w_a$ are constants, and possible deviations of $(w_0, w_a)$ from $(-1, 0)$ would signify departure from the $\Lambda$CDM model.

### 2.2.3. Mock Data

The mock data employed in our study are simulated based on the matter power spectrum derived from the ds14-a set of Dark Sky $N$-body simulation series (Skillman et al. 2014). The simulations were carried out using fiducial values for the cosmological parameters as follows: $\{\Omega_m, \Omega_b, \Omega_\Lambda, h, n_s, \sigma_8\} = \{0.295, 0.0468, 0.705, 0.688, 0.9676, 0.8344\}$. This mock data will be used in the likelihood analysis outlined in Section 3 to explore how inaccuracies in nonlinear prescriptions can lead to biases in parameter estimation.

We should note that this work only deals with the impact of the choice of nonlinear prescription on the matter power spectrum and baryonic processes are ignored. However, at small scales, baryonic feedback also plays an important role, and accurate cosmological inferences require proper modeling of these effects. Therefore, for a thorough analysis based on the matter power spectrum at large $k$, one needs to properly incorporate the baryonic effects. This could be done, e.g., through using $N$-body simulations with baryons (Schaye et al. 2010; Le Brun et al. 2014; McCarthy et al. 2016; Springel et al. 2017), using halo models fitted to $N$-body simulations with proper baryonic add-ons (e.g., Semboloni et al. 2011; van Daalen et al. 2011; Casarini et al. 2012; Castro et al. 2018; Debackere et al. 2019; Huang et al. 2019; Schneider et al. 2019; Martinet et al. 2021; Mead et al. 2021; Salcido et al. 2023), or by using the version of the EFTofLSS with baryon impact included (Bragança et al. 2021). We leave this analysis to future work.

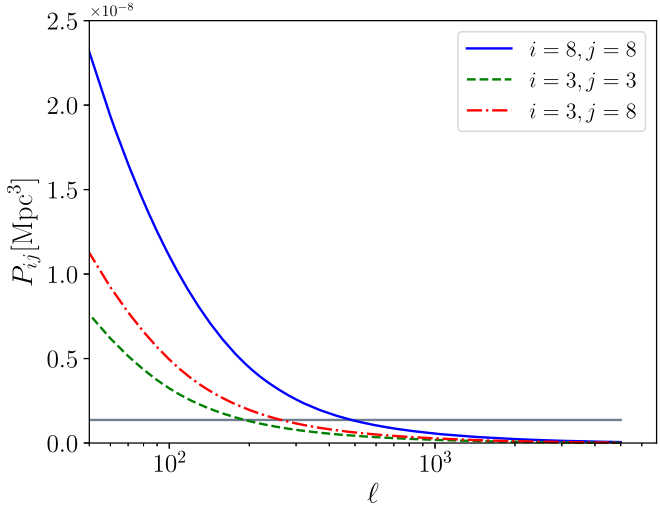

**Figure 3.** The cosmic shear power spectrum $P_{ij}(\ell)$ for different bin pairs with $k_{\max} = 2h\,\mathrm{Mpc}^{-1}$.

### 2.3. Nonlinear Cutoffs

Given that cosmic shear is sensitive to small scales (up to $k \sim 7h\,\mathrm{Mpc}^{-1}$; Taylor et al. 2018b), the challenge of modeling nonlinearities at these scales can lead to inaccuracies in the matter power spectrum. To overcome this problem, we introduce a cutoff to exclude scales with poorly modeled nonlinear behavior. For this purpose, we use a Gaussian window function that smoothly attenuates the contribution of the highly nonlinear scales with $k > k_{\max}(z)$. This choice would circumvent numerical issues caused by a sharp window function and prevents the introduction of numerical noise into our calculations. We conservatively choose $k_{\max}(0) = 2h\,\mathrm{Mpc}^{-1}$ and allow for its redshift dependence through (Sprenger et al. 2019)

$$k_{\max}(z) = k_{\max}(0) \times (1+z)^{2/(2+n_s)}. \quad (9)$$

This is to acknowledge that at higher redshifts, with shorter time available for the evolution of structures, a greater proportion of scales remains in the linear regime. For the sake of comparison, we also use $k_{\max}(0) = 0.5h\,\mathrm{Mpc}^{-1}$ in a few cases.

Figure 3 displays the shear power spectrum for various bin pairs with $k_{\max} = 2h\,\mathrm{Mpc}^{-1}$. The gray solid horizontal line presents the level of shot noise (Equation (6)). We see that noise starts to dominate over the shear power spectrum at larger scales as the redshift decreases. This is because of the relatively low amplitude of $P_{ij}(\ell)$ at these redshifts due to the shorter distances of the sources to the observer and therefore being affected by fewer intervening structures, thus leading to a smaller signal.

Moreover, since cosmic shear observations are commonly reported as angular power spectra, a multipole cutoff for the highly nonlinear scales, $\ell_{\max}$, is often used as a straightforward choice. In this work we compare the results for two choices of $\ell_{\max}$, i.e., 1500 and 5000.

Figure 4 shows the sensitivity of the shear power spectrum for the bin $i = 5$ to different cosmological parameters, for two different $k_{\max}$. We see that the highest sensitivity to most of these parameters is at $\ell \lesssim 2000$, while the contribution from higher $\ell$'s is nonnegligible, in particular for the larger $k_{\max}$.

It is important to note that the mapping of the cutoff from $k$ space (i.e., $k_{\max}$) to the multipole space (i.e., $\ell_{\max}$) is nontrivial as each multipole $\ell$ encompasses contributions from different $k$ in a redshift-dependent way (through Limber approximation $k \sim \ell/r(z)$). This is due to the inherent nature of lensing as a projected effect across a potentially broad redshift range. Therefore, we also use an alternative truncation approach in $\ell$ space, where for an assumed $k_{\max}$, the maximum multipole used in the analysis is allowed to depend on the redshift of the bin. This method was also outlined in Martinelli et al. (2020). The results of the analysis with this truncation algorithm are presented in Section 4, and we refer to it as the varying-$\ell_{\max}$ or $\ell_{\mathrm{var}}$ cutoff.

It should be noted that the various methods of discarding poorly modeled nonlinear scales would be a serious challenge when it comes to the analysis of true data. Accurate discrimination of the contribution from different redshifts to a single observed bin requires the application of specialized inversion methods (e.g., see Bernardeau et al. 2014; Taylor et al. 2018a).

### 3. Analysis

Here we use two approaches to investigate the sensitivity of cosmological inferences to the nonlinear prescription. First we modify CosmoMC[1] (Lewis & Bridle 2002; Lewis 2013), which is a publicly available code for sampling from the parameter space in cosmology to calculate the cosmic shear likelihood for a Euclid-like survey and for different cosmologies. CosmoMC uses HF for modeling the MNL regime. As the second approach, we perform Fisher analysis for both HF and EFT and compare their capability in estimating errors of the cosmological parameters.

#### 3.1. Cosmic Shear Likelihood

We assume that the data points (here, the $a_{\ell m}$) have a Gaussian distribution with their covariance matrix for the fiducial cosmology modeled by

$$C_{ij}^{\mathrm{obs}}(\ell) = P_{ij}^{\mathrm{fid}}(\ell) + N_{ij} \quad (10)$$

(see Equations (5) and (6)). We then have (Audren et al. 2013; Sprenger et al. 2019)

$$-2\ln\mathcal{L} = \sum_\ell (2\ell+1)f_{\mathrm{sky}}\left(\frac{d^{\mathrm{mix}}(\ell)}{d^{\mathrm{th}}(\ell)} + \ln\frac{d^{\mathrm{th}}(\ell)}{d^{\mathrm{obs}}(\ell)} - N_{\mathrm{bin}}\right), \quad (11)$$

where $\mathcal{L}$ is the likelihood of the (simulated) data given the theoretical model, and $f_{\mathrm{sky}}$ is the observed fraction of sky. The $d$ represents the determinants of the covariance matrices,

$$d^{\mathrm{th}}(\ell) = \det(P_{ij}^{\mathrm{th}}(\ell) + N_{ij}), \quad (12)$$

$$d^{\mathrm{obs}}(\ell) = \det(P_{ij}^{\mathrm{fid}}(\ell) + N_{ij}), \quad (13)$$

$$d^{\mathrm{mix}}(\ell) = \sum_k \det\left(N_{ij} + \begin{cases} P_{ij}^{\mathrm{th}}(\ell) & j \neq k \\ P_{ij}^{\mathrm{fid}}(\ell) & j = k \end{cases}\right), \quad (14)$$

where $P_{ij}^{\mathrm{th}}(\ell)$ and $P_{ij}^{\mathrm{fid}}(\ell)$ are the theoretical and fiducial power spectra.

---

[1] https://cosmologist.info/cosmomc/

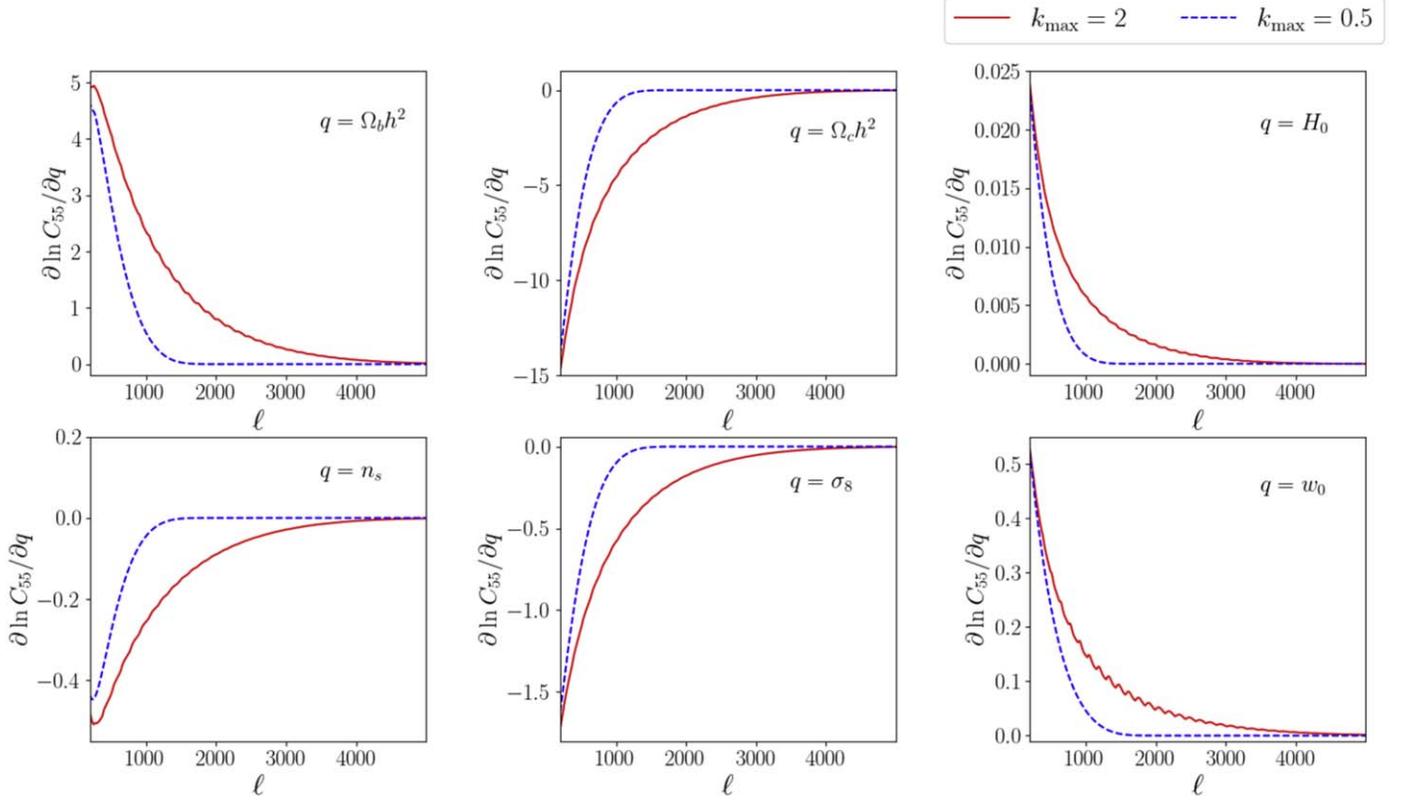

**Figure 4.** Sensitivity of the shear power spectrum (including the noise contribution, Equation (5)) to different cosmological parameters, for the intermediate redshift bin $i = 5$.

### 3.2. Fisher Formalism

In this section, we employ the Fisher matrix formalism to estimate the errors of the cosmological parameters. Using Bayes's theorem, the posterior distribution $p(\boldsymbol{q}|\boldsymbol{D})$ of the model parameters $\boldsymbol{q}$, given the data $\boldsymbol{D}$, is

$$p(\boldsymbol{q}|\boldsymbol{D}) = \frac{\mathcal{L}(\boldsymbol{D}|\boldsymbol{q})p(\boldsymbol{q})}{p(\boldsymbol{D})}, \quad (15)$$

where $p(\boldsymbol{q})$ describes the prior information on our parameters and $p(\boldsymbol{D})$ is the evidence. The Fisher matrix is then defined as

$$F_{\alpha\beta} \equiv \left\langle -\frac{\partial^2 \ln \mathcal{L}}{\partial q_\alpha \partial q_\beta}\bigg|_{\boldsymbol{q}_{\rm fid}} \right\rangle, \quad (16)$$

where $\alpha$ and $\beta$ label the parameters of interest and the derivatives are evaluated at the fiducial parameter values $\boldsymbol{q}_{\rm fid}$ (Vogeley & Szalay 1996; Tegmark et al. 1998). The Fisher matrix is a measure of the curvature of the logarithm of the likelihood surface. In case of Gaussian parameter distribution, the parameter covariance matrix is given by the Fisher inverse.

For weak lensing, the Fisher matrix can be obtained by summing over contributions from all pairs of observed redshift bins and is given by

$$F_{\alpha\beta} = f_{\rm sky} \sum_\ell \frac{(2\ell+1)}{2} \sum_{ijkl} \frac{\partial C_{ij}(\ell)}{\partial q_\alpha} C_{jk}^{-1}(\ell) \frac{\partial C_{kl}(\ell)}{\partial q_\beta} C_{li}^{-1}(\ell)$$

$$= f_{\rm sky} \sum_\ell \frac{(2\ell+1)}{2} {\rm Tr}\left[\frac{\partial C(\ell)}{\partial q_\alpha} C^{-1}(\ell) \frac{\partial C(\ell)}{\partial q_\beta} C^{-1}(\ell)\right], \quad (17)$$

where $C$ is the covariance matrix of the fiducial cosmology (Blanchard et al. 2020).

In the process of estimating the Fisher matrix, it is important to carefully consider the step size used for taking derivatives. Choosing an appropriate step size is crucial for obtaining stable and reliable results. Too-large steps can lead to inaccurate estimates of the derivatives and introduce significant errors in the result. On the other hand, a too-small step size would result in numerical instabilities. Finding an optimal step size involves a balance between accuracy and stability. We identified the proper step size by performing numerical tests and assessing the reliability of the results. We tried to begin with a smart initial guess, usually of the order of the expected $1\sigma$ uncertainty, and gradually decreased the step size until the derivatives no longer change. Often, acceptable convergence is reached before numerical noise contribution gets significant.

### 4. Results

In this section we present the results on the sensitivity of the cosmological inferences from the cosmic shear probe of a Euclid-like experiment to the two nonlinear prescriptions used in this work and the various nonlinear cutoffs applied.

We first quantify the contribution of each multipole at different redshift bins to the measurement of various parameters. For this purpose we define

$$\Gamma_{i,\alpha}(\ell) = A_i(\ell) \frac{2\ell+1}{2} \sum_{jkl} \frac{\partial C_{ij}(\ell)}{\partial q_\alpha} C_{jk}^{-1}(\ell) \frac{\partial C_{kl}(\ell)}{\partial q_\alpha} C_{li}^{-1}(\ell), \quad (18)$$

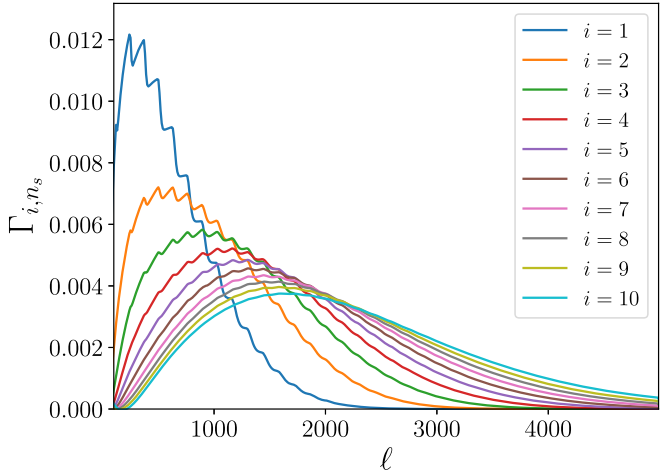

**Figure 5.** The contribution of various redshift bins to the measurement of $n_s$ as a function of $\ell$ (Equation (18)). The overall behavior remains largely unchanged for other parameters.

**Table 3**
The Estimated 1$\sigma$ Uncertainty of Cosmological Parameters in the $\Lambda$CDM Scenario from Cosmic Shear for

| | $\Lambda$CDM | | | | | |
|---|---|---|---|---|---|---|
| | Nonlinear Prescription | $\Omega_b h^2$ | $\Omega_c h^2$ | $H_0$ | $n_s$ | $\sigma_8$ |
| $\ell = 1500$ | EFT | 0.01 | 0.03 | 10.0 | 0.05 | 0.2 |
| | HF | 0.01 | 0.02 | 7.1 | 0.02 | 0.09 |
| $\ell = 5000$ | EFT | 0.008 | 0.009 | 3.4 | 0.02 | 0.08 |
| | HF | 0.007 | 0.01 | 4.5 | 0.02 | 0.06 |

**Note.** The predictions are compared for EFT and HF.

**Table 4**
Similar to Table 3 but for the $w$CDM Scenario

| | $w$CDM | | | | | | |
|---|---|---|---|---|---|---|---|
| | Nonlinear Prescription | $\Omega_b h^2$ | $\Omega_c h^2$ | $H_0$ | $n_s$ | $\sigma_8$ | $w$ |
| $\ell = 1500$ | EFT | 0.01 | 0.03 | 10.2 | 0.05 | 0.2 | 0.05 |
| | HF | 0.01 | 0.02 | 7.1 | 0.03 | 0.09 | 0.05 |
| $\ell = 5000$ | EFT | 0.008 | 0.01 | 3.9 | 0.02 | 0.1 | 0.03 |
| | HF | 0.009 | 0.01 | 5.9 | 0.02 | 0.07 | 0.02 |

**Note.** We used the absolute value of $w$ in the relative error estimation.

**Table 5**
Similar to Table 4 but for the $(w_0, w_a)$ CDM Scenario

| | $(w_0, w_a)$ CDM | | | | | | | |
|---|---|---|---|---|---|---|---|---|
| | Nonlinear Prescription | $\Omega_b h^2$ | $\Omega_c h^2$ | $H_0$ | $n_s$ | $\sigma_8$ | $w_0$ | $w_a$ |
| $\ell = 1500$ | EFT | 0.01 | 0.03 | 12.2 | 0.05 | 0.2 | 0.1 | 0.8 |
| | HF | 0.01 | 0.05 | 16.6 | 0.09 | 0.2 | 0.1 | 0.7 |
| $\ell = 5000$ | EFT | 0.008 | 0.01 | 4.4 | 0.03 | 0.1 | 0.07 | 0.6 |
| | HF | 0.01 | 0.03 | 10.4 | 0.06 | 0.1 | 0.07 | 0.4 |

**Table 6**
Marginalized 1$\sigma$ Errors and Biases for the Cosmological Parameters for the Three Scenarios $\Lambda$CDM, $w$CDM, and $(w_0, w_a)$CDM with $k_{max} = 2h$ Mpc$^{-1}$

| | | $\Lambda$CDM | | $w$CDM | | $(w_0, w_a)$CDM | |
|---|---|---|---|---|---|---|---|
| $q$ | $\ell_{max}$ | $\sigma$ | $B$ | $\sigma$ | $B$ | $\sigma$ | $B$ |
| $\Omega_b h^2$ | 1500 | 0.012 | 1.9 | 0.011 | 0.7 | 0.013 | 1.0 |
| | 5000 | 0.006 | 1.4 | 0.010 | 1.9 | 0.011 | 1.7 |
| | $\ell_{var}$ | 0.005 | 0.4 | 0.006 | 0.2 | 0.009 | 0.2 |
| $\Omega_c h^2$ | 1500 | 0.015 | 1.05 | 0.015 | 0.3 | 0.024 | 0.6 |
| | 5000 | 0.007 | 0.4 | 0.013 | 0.8 | 0.015 | 1.3 |
| | $\ell_{var}$ | 0.005 | 1.2 | 0.011 | 0.9 | 0.017 | 0.3 |
| $H_0$ | 1500 | 5.7 | 1.6 | 6.2 | 0.05 | 8.5 | 0.8 |
| | 5000 | 3.01 | 0.7 | 5.2 | 1.5 | 5.6 | 1.8 |
| | $\ell_{var}$ | 1.2 | 0.3 | 3.8 | 0.4 | 6.2 | 0.7 |
| $n_s$ | 1500 | 0.01 | 4.0 | 0.02 | 3.9 | 0.03 | 0.3 |
| | 5000 | 0.01 | 5.2 | 0.01 | 4.0 | 0.02 | 1.1 |
| | $\ell_{var}$ | 0.04 | 0.7 | 0.04 | 0.8 | 0.06 | 0.2 |
| $\sigma_8$ | 1500 | 0.0039 | 2.8 | 0.0054 | 0.7 | 0.0094 | 1.6 |
| | 5000 | 0.0033 | 3.7 | 0.0036 | 3.7 | 0.0060 | 3.3 |
| | $\ell_{var}$ | 0.0047 | 3.1 | 0.0049 | 2.9 | 0.0142 | 2.4 |
| $w_0$ | 1500 | ⋯ | ⋯ | 0.04 | 1.9 | 0.06 | 0.2 |
| | 5000 | ⋯ | ⋯ | 0.01 | 1.0 | 0.03 | 1.7 |
| | $\ell_{var}$ | ⋯ | ⋯ | 0.06 | 0.7 | 0.11 | 1.0 |
| $w_a$ | 1500 | ⋯ | ⋯ | ⋯ | ⋯ | 0.22 | 1.6 |
| | 5000 | ⋯ | ⋯ | ⋯ | ⋯ | 0.18 | 1.6 |
| | $\ell_{var}$ | ⋯ | ⋯ | ⋯ | ⋯ | 0.40 | 1.5 |

**Note.** In most cases we observe that the bias increases as we transition from $\ell_{max} = 1500$ to $\ell_{max} = 5000$ and biases decrease in the $\ell_{var}$ case.

where $A_i$ is chosen so that $\sum_\ell \text{Tr}\, \Gamma_{i,\alpha}(\ell) = 1$. As an example, Figure 5 provides a visual representation of $\Gamma_{i,n_s}(\ell)$. The overall patterns are quite similar for other parameters. We see that the peak of the bin contributions moves to higher $\ell$ as the bin number (i.e., redshift) increases. This is expected, since for a given physical scale, larger distances correspond to smaller angular scales, leading to higher contributions for larger multipoles.

Now we present the results of the Fisher forecast (Section 3.2) applied to the various simulated data sets with different modelings of the MNL regime (EFT and HF) and for different cosmological scenarios ($\Lambda$CDM, $w$CDM, and $(w_0, w_a)$CDM). Tables 3–5 compare the forecasted errors with different $\ell_{max}$ as nonlinear cuts. The EFT parameter, $c_s$, is assumed to be fixed. For a more thorough analysis, one should marginalize over these sound speeds (e.g., see Safi & Farhang 2021). However, the degeneracy between the $c_s$ and the standard cosmological parameters turns out to be small for most cases considered here, leaving the forecasted errors on cosmological parameters quite unchanged. On the other hand, for certain choices of nonlinear scales and fiducial values of sound speeds, we find numerical instability issues. We therefore only report the results with fixed $c_s$ and ignore possible small increases in the errors of the standard parameters due to the marginalization process.

*Parameter Bias.* For a quantitative analysis of the impact of the MNL regime of structures on the parameter estimation, we define the bias parameter $B_q = (\bar{q} - q_{fid})/\sigma_q$, where $\bar{q}$ and $\sigma_q$ correspond to the mean and estimated error of the parameter $q$, as found from the likelihood analysis (Section 3.1). Marginalized 1D parameter likelihoods are presented in Figure 6. The values of parameter biases and 1$\sigma$ errors are listed in Table 6.

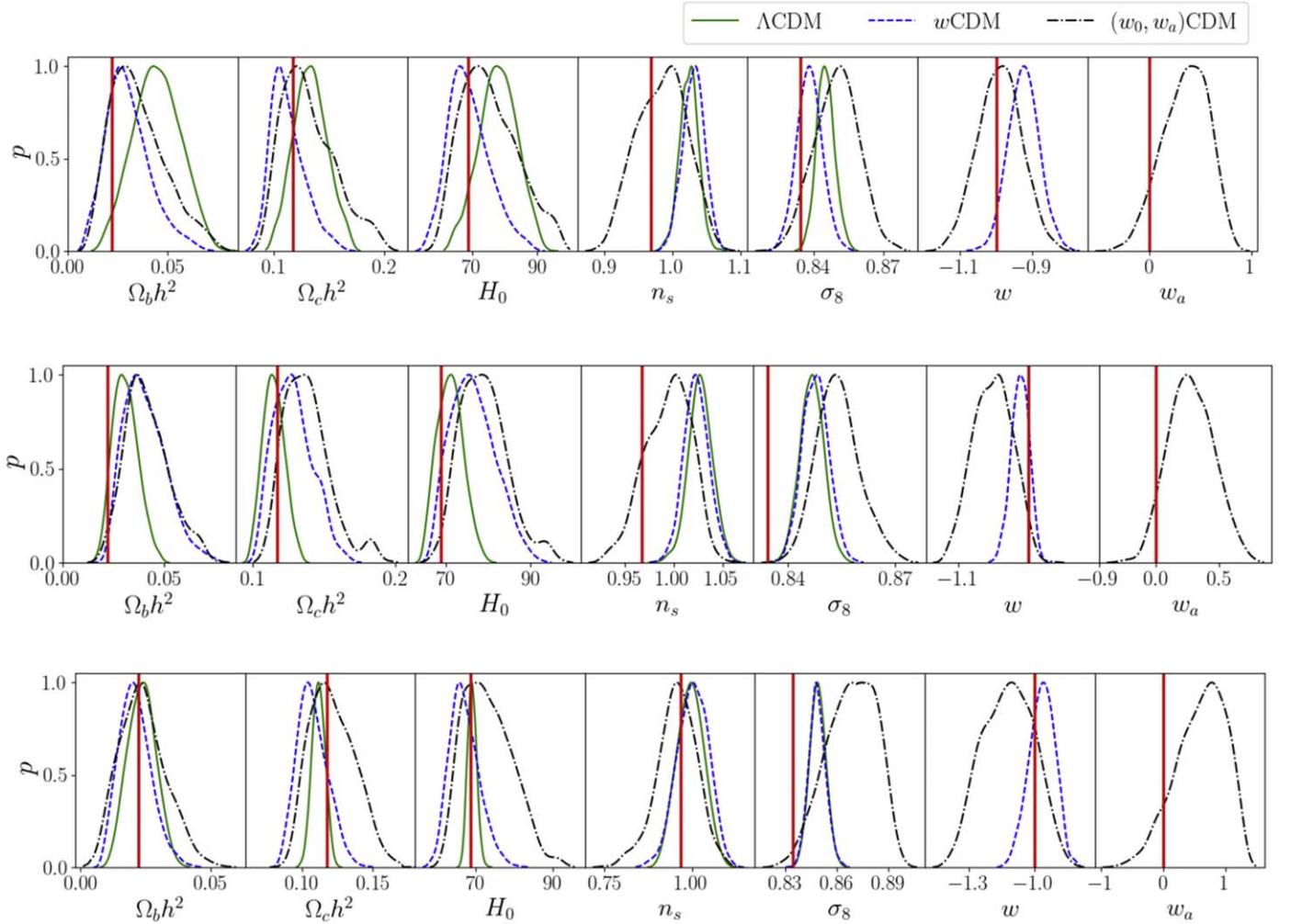

**Figure 6.** Marginalized 1D parameter likelihoods with cosmic shear measurements for a Euclid-like experiment for three different cosmological scenarios. The panels correspond to different nonlinear cutoffs: $\ell_{max} = 1500$ (top), $\ell_{max} = 5000$ (middle), and varying $\ell_{max}$ with $k_{max} = 0.25h$ Mpc$^{-1}$ (bottom).

In most cases we observe an increase in parameter bias when transitioning from $\ell = 1500$ to $\ell = 5000$. This is expected as the inclusion of more nonlinear scales in the analysis would possibly lead to larger deviation of the measured mean values. Also the errors tend to decrease when incorporating smaller scales (i.e., more information). Therefore, the parameter biases are enhanced.

*Figure of Merit*. As a quantitative assessment of the overall power of an experiment to constrain the volume of the parameter space in a given cosmological framework, we use the following figure of merit (FoM; Albrecht et al. 2006):

$$\text{FoM} = -\frac{1}{2} \ln(\det \boldsymbol{C}), \quad (19)$$

where $\boldsymbol{C}$ is the parameter covariance matrix. As constraints on cosmological parameters become tighter, the volume of the associated parameter space shrinks, resulting in a reduced determinant of $\boldsymbol{C}$. This, in turn, leads to a higher FoM. Table 7 provides an illustration of the FoMs within the marginalized $(w_0, w_a)$ parameter space with $\ell_{max} = 1500$ and $\ell_{max} = 5000$. These computations encompass both nonlinear prescriptions (HF and EFT) and are conducted with two different values of $k_{max}$, specifically, $k_{max} = 0.5h$ Mpc$^{-1}$ and $k_{max} = 2h$ Mpc$^{-1}$. It is evident from the table that with higher $k_{max}$ and $\ell_{max}$, FoM increases due to the greater amount of information from small-scale modes being exploited in the analysis. This, again, highlights the significance of the proper modeling of the MNL regime as inaccurate modeling would lead to biased inferences.

**Table 7**
FoM Calculated with HF and EFT Nonlinear Models for $k_{max} = 0.5h$ Mpc$^{-1}$ and $k_{max} = 2h$ Mpc$^{-1}$

| | EFT | | HF | |
|---|---|---|---|---|
| $\ell_{max}$ | $k_{max} = 0.5$ | $k_{max} = 2$ | $k_{max} = 0.5$ | $k_{max} = 2$ |
| 1500 | 2.31 | 3.15 | 2.29 | 3.06 |
| 5000 | 2.33 | 3.96 | 2.41 | 4.33 |

## 5. Conclusion

Future large-scale structure surveys are going to explore the MNL regime with unprecedented precision. This enhanced precision highlights the significance of the choice for the nonlinear theoretical framework as errors in the modeling process may potentially introduce biases in parameter estimation. The main objective of this study was to explore how the choice of nonlinear prescription would impact the ability of a Euclid-like survey for accurate and precise estimation of cosmological parameters from cosmic shear measurements.

To quantify the precision of the measurements, we performed a Markov Chain Monte Carlo analysis and explored the likelihood surface of the parameters of interest to measure the induced biases in parameter estimation. We find, in most cases, biases tend to increase as we transition from $\ell_{\max} = 1500$ to $\ell_{\max} = 5000$. This outcome is expected because of the greater amount of information delivered with the inclusion of higher multipoles. We also observe that in most cases, when varying $\ell_{\max}$ (as shown in the $\ell_{\mathrm{var}}$ case in Table 6), there is less parameter bias compared to the other cases. In certain cosmological parameters, this reduction is attributed to the enhanced precision in $\ell_{\mathrm{var}}$. To explore the effect of nonlinear modeling on the accuracy of parameter measurements, we also applied the Fisher matrix formalism to the three dark energy models used in this work. We computed the marginalized FoM for dark energy parameters as well.

In brief, the choice of nonlinear methods can substantially impact the accuracy and precision of parameter estimation and may introduce nonnegligible biases. This underscores the necessity for further improvements in the way the future large-scale data are analyzed and interpreted.